\newcounter{myctr}
\def\myitem{\refstepcounter{myctr}\bibfont\noindent\ifnum\themyctr>9\else\phantom{0}\fi\hangindent17pt\themyctr.\enskip}

\documentclass[12pt]{article}
\usepackage{amsmath}
\usepackage{tikz}
\usepackage[compat=1.1.0]{tikz-feynman}
\usepackage{amssymb}
\usepackage{amscd}
\usepackage{dsfont}
\usepackage{verbatim}
\usepackage{color}
\numberwithin{equation}{section}
 
\usepackage{amsfonts}
\usepackage{amssymb}
\usepackage{textcomp}
\everymath{\displaystyle}
\def\be{\begin{equation}}
\def\ee{\end{equation}}
\def\beq{\begin{eqnarray}}
\def\eeq{\end{eqnarray}}
\def\bn{\begin{eqnarray*}}
\def\en{\end{eqnarray*}}

\def\a{\alpha}
\def\b{\beta}

\def\g{\gamma}

\def\pd{\partial}
\def\e{\epsilon}

\def\l{\lambda}

\def\cL{{\cal{L}}}

\def\cO{{\cal{O}}}

\begin{document}




\title{Infrared Effects and the Soft Photon Theorem in Massive QED}

\author{T R Govindarajan and Nikhil Kalyanapuram}

\maketitle

\begin{center}
Chennai Mathematical Institute, Kelambakkam\\
Siruseri, Tamil Nadu 600113,
India\\
trg@cmi.ac.in, trg@imsc.res.in\\
nikhilkaly@cmi.ac.in
\end{center}

\begin{abstract}
Stuckelberg QED with massive photon is known to be
renormalizable. But the limit of the mass going to zero is
interesting because it brings the resolution to infrared questions
through the role of Stuckelberg field at null infinity in addition
to providing new asymptotic symmetries. Such symmetries facilitate
the soft photon theorems also.
\end{abstract}

\section{Introduction}	
Asymptotic symmetries depend on the exact masslessness
of the photon. But the current bound on the Compton wavelength of
photon is that of solar system \cite{PDB,Erwin}. 
At no stage we may be able to
conclude operationally exact masslessness of the photon. Given this
scenario it is more appropriate to study the massive photon QED
theory. But gauge invariance is crucial for studying interactions.
Such a massive QED with gauge invariance is Stuckelberg theory. This
can be considered as a limit of Higgs model 
where phase of the Higgs alone survives. In this model the infrared
questions can also be regulated. 
The extra degree of freedom play an important role providing 
sky photons. We study the soft photon theorem
which is another manifestation of the asymptotic 
symmetry and its consequences.

The soft photon and soft graviton theorems, described by Weinberg
\cite{Wein1,Wein3} are rules for the modification of transition rates
when an investigation is carried out to deal with the infrared
divergences which arise due to the exact masslessness of these 
particles. Referring to the soft photon theorem 
in particular, infrared
divergences manifest themselves in the virtual transmission of very
soft photons between external particle legs. By taking into account
all such communications and then proceeding to include the effects
of all possible real soft photon emissions, while maintaining a cut
off that corresponds to some energy resolution, a cancellation of
the infrared cutoff is achieved, thereby supplying the soft photon
theorem. The actual statement of the theorem is as follows. If one
considers a reaction $\alpha\rightarrow\beta$ involving photons,
the reaction rate has to be modified to account for the possible.
emissions of soft photons having total energy less than or equal to
a minimum energy resolution $\Delta E$. The modification takes
the form:
\be
\Gamma_{\beta\alpha} = \left(\frac{\Delta E}{\Lambda}\right)^{A}b(A)\Gamma^{0}_{\beta\alpha}
\ee
Here, $\Gamma^0$ refers to the bare reaction rate which does not
refer to any soft photons. $\Lambda$ is a renormalization scale.
The functions $A$ and $b(x)$ are defined in the book of Weinberg
\cite{Wein1} Eqs.2.16 and 2.50 respectively.

While attempting the calculation of a Feynman diagram that is
characterised by such divergences (for example, the electron
self energy contribution or the vertex), the problem can be
temporarily circumvented by appealing to a small mass parameter
$m_\gamma$. This parameter serves as a convenient regulator of the
integrals, which despite its appeal, violates gauge invariance.
Hence, an approach which offers both the technical ease of a photon
mass which nonetheless respects gauge invariance would be
extremely desirable.

These qualitative points being made, it may be noted that
originally, the infrared divergences were regarded as essentially
unphysical, which ultimately had no bearing on the final answers,
so long as one took into account the possibility of soft photon
emissions, keeping in mind the right physical picture. In recent
years however, there has been a resurgence of interest in soft
photon theorems and their relationship to asymptotic symmetries
and the Ward identities, described in 
Strominger\cite{Strom1,Strom2,Strom3,Strom4}. Following 
there are several interesting questions that have been 
posed in understanding BMS symmetries in gravity\cite{ALOK}, 
effective action in QFT's and gauge theories 
including non abelian gauge theories.
Balachandran et al\cite{BALQED} have pointed out 
the exact massless photon QED may violate Lorentz 
symmetry. Similar analysis also leads to color symmetry being broken
\cite{BALQCD}.    
To note one point of interest, the classical Lienard-Wiechert 
potential displays a discontinuity at infinity, 
which when investigated, reveals a family of conserved 
quantities in QED.

Having noted these aspects of the infrared regime of QED, it may be
interesting to consider what we 
obtain when we introduce a very tiny photon
mass in a gauge invariant fashion and proceed to probe the
infrared regime of the theory. To do this, recourse may be had to
the Stuckelberg mechanism.

The Stuckelberg mechanism, simply put, is an abelian Higgs
mechanism. It involves the introduction of an auxiliary scalar
field, supplying the photon a small mass $m_\gamma$, which arises as
a coupling constant between the photon and the Stuckelberg
particle. Once we introduce this mechanism, we will study
the analogue of the soft photon theorems in this theory. It makes
sense then, to speak of the infrared regime since the photon mass,
of $\leq ~ \cO({10^{-18}}) eV$ constrained by experiments 
\cite{PDB,Photon,Photon2}, may be treated as extremely small. 
The actual experimental
inaccuracy may in fact be much higher than this limit.

The goal of this paper is twofold. Firstly, we will investigate the
soft photon theorem for a massive photon. In doing so, naturally,
the infrared cutoff will be taken as the photon mass, now suitably
regulated in a gauge invariant fashion. When real and virtual
transitions are taken into account and the limit
$m_\gamma \rightarrow 0$ is attempted, a divergence is encountered,
the vanishing of which will be tied up with charge conservation. We
unambiguously obtain the original result then, this time with no
reference to an arbitrary minimum wavelength, the theory now being
effectively regulated by the photon mass parameter. In principle 
$\cO(m^2)$ corrections would be there which will be negligible.

Secondly, the classical Lienard-Wiechert potentials for 
the massive photon field is investigated, with 
the aim of recovering the correct
antipodal matching points in a suitable limit. In doing so, we
obtain another regulating parameter, this time, encoding the
relationship between the bound of the photon mass and the limit of
the radius going to infinity. Conditions on the photon mass are
obtained once the antipodal matching condition is identified and
the implications of this on the regulating parameter are
investigated.

We briefly introduce the Stuckelberg model in Sec 2. In Sec 3. 
we explore infrared divergences and soft theorems, In Sec 4, we 
consider modified Lienard Wiechart potential relevant for our purpose
and consider asymptotic limits. In Sec.5 we present our conclusions
taking into account several developements in the recent revisit 
of this question and some details of future works and publications.  

\section{\bf Review of Stuckelberg theory}
Stuckelberg theory is defined by a modified Maxwell/Proca lagrangian
given by:
\begin{equation}
\mathcal{L}_S = -\frac{1}{4}F_{\mu\nu}F^{\mu\nu} + 
\frac{m_\gamma^2}{2}\left(A_\mu + 
\frac{1}{m_\gamma}\partial_\mu S\right)
\left(A^\mu + \frac{1}{m_\gamma}\partial^\mu S\right).
\end{equation}

\noindent The lagrangian is invariant under the gauge transformations:
\be A_\mu~\longrightarrow ~A_\mu~-\pd_\mu\l, \qquad S~\longrightarrow
S~-~m_\g~\l
\ee 
where $\l$ is the gauge parameter. In this theory, 
the Lorenz gauge is generalised to the mass shell condition on $S$.

Stuckelberg theory is known to be ultraviolet renormalizable for 
all values of the mass paramter $m_\g$ as explained in the review 
paper by Ruegge and Altaba\cite{Ruegg}. The infrared divergence 
when $m_\g \longrightarrow 0$ would be of interest.
 
There are two avenues which may be pursued if one wishes to study 
the infrared effects that arise in Stuckelberg theory. One may 
study the photon-Stuckelberg vertex and include its effects in 
radiative corrections. This procedure however is cumbersome and not 
very illuminating. Instead, a suitable gauge transformation will 
make the analysis simpler. By effecting the gauge transformation 
\be A_\mu~\longrightarrow A_\mu ~+~ \frac{1}{m_\g}\pd_\mu S 
~=~ \widetilde{A_\mu},
\ee 
the covariant derivative is modified to,

\begin{equation}
    \partial_\mu ~-~ ie\widetilde{A}_{\mu} + 
\frac{ie}{m_\g}\partial_\mu S.
\end{equation}

This is just a convenient writing of the covariant derivative. 
It shows that the gauge part of the derivative coming from 
transformations of the matter field are eliminated by the 
corresponding transformations of the Stuckelberg field.

We study scalar QED to avoid technical complications of spinorial
polarisations. The additional term in the Lagrangian is
\be \cL_\phi~=~(D_\mu\varphi)^*~(D^\mu\varphi)~+~m^2 \varphi^*\varphi
\ee

The relevant interaction term for scalar QED may then be written as,
\be
\mathcal{H}_{I} = ieA_{\mu}\left(\varphi^*\pd^{\mu}\varphi - 
\partial^{\mu}\varphi^*\varphi\right)~-~ 
\frac{i~e}{m_\g}\pd_\mu S\left(\varphi^*\pd^{\mu}\varphi~-~ 
\varphi\pd^{\mu}\varphi^*\right).
\ee

Now, the photon scalar and the Stuckelberg scalar vertices 
read as follows.

\begin{figure}[ht!]
\centering
\includegraphics[width=0.9\textwidth]{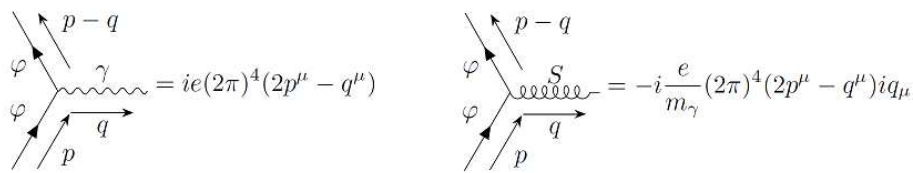}
\caption{\label{fig:Diagrams}Feynman Rules for Stuckelberg Theory.}
\end{figure}

The propagators are,
\be
\frac{-i}{(2\pi)^4}\frac{\left(g_{\mu\nu} + \frac{q_\mu q_\nu}{m_\g^2}
\right)}{q^2 + m^2 - i\e}
\ee
for the photon and,
\be
\frac{-i}{(2\pi)^4}\frac{1}{q^2 + m^2 - i\e}
\ee
for the Stuckelberg field.

As pointed earlier, renormalizability of the Stuckelberg scalar 
QED can be demonstrated by constructing nilpotent BRST operator
\cite{Ruegg}. But we will focus on questions of 
infrared behaviour. Since there 
is mass for the photon there is no infrared divergence however 
small the mass is. We would like to analyse 
$m_\g ~\rightarrow~0$ limit where potentially it can appear. 

\section{\bf Infrared limit and soft theorems}
With the Feynman rules in hand, the calculation of the soft 
factors proceeds in much the same way as that of Weinberg. Consider 
an arbitrary Feynman diagrams with legs labelled as either incoming 
or outgoing. Defining the symbol $\eta$, which takes values $+1$ 
for outgoing and $-1$ for incoming legs, we may proceed to compute 
the soft factors. 

First, we should qualify the meaning of the term soft factor. A soft factor is a correction introduced into a matrix element due to the addition of an outgoing photon leg onto any of the external charged particle legs of the amplitude in question. This addition of this so called soft photon leg is quantified by the multiplication by the charged particle-photon vertex and the inclusion of a propagator of the form $((p+\eta q) + m^2 - i\epsilon)^{-1}$, where $p$ is the momentum of the particle leg and $q$ is the photon (or Stuckelberg) momentum, which is very close to zero. Due to the smallness of $q$, $p$ is essentially on the mass shell. $q^2$ may be neglected. These leaves behind $(2p\cdot q\eta - i\epsilon)^{-1}$. This has to now be multiplied by a suitable vertex factor.

The photon soft factor remains unchanged,

\be
S^{1}_\gamma(q) =  e\frac{\eta p^{\mu}}{p\cdot q - i\eta\epsilon}.
\ee

The Stuckelberg soft factor takes the form,
\be
S^{1}_S(q) =  -i\frac{e}{m_{\g}}\frac{\eta p\cdot q}{p\cdot q 
-i\eta\e} = -i\eta\frac{e}{m_\g}.
\ee

It is clear from the form of these soft factors that the 
factorization necessary to sum over all possible virtual 
or real emissions of soft particles is ensured 
and this holds true in the new theory as well.
The factors that correct the $S$-matrix that correspond to a 
delivery of one soft particle from one leg to another is expressed 
by the following for a photon.
\be
A_{\g}(q) = \frac{-i}{(2\pi)^4}\frac{\left(g_{\mu\nu} + 
\frac{q_\mu q_\nu}{m_\g^2}\right)}{q^2 + m^2 - i\epsilon}
\sum_{ij}\frac{e_ie_j\eta_i\eta_jp^\mu_ip^\nu_j}
{\left(p_i\cdot q - i\eta_i\epsilon\right)
\left(-p_j\cdot q - i\eta_j\epsilon\right)}
\ee
and by the following sum for a Stuckelberg particle,
\be
A_{S}(q) = \frac{-i}{(2\pi)^4}\frac{1}{q^2 + m^2 - i\e}\frac{1}{m_\g^2}
\sum_{ij}\frac{e_ie_j\eta_i\eta_j(p_i\cdot q) (p_j\cdot q)}
{\left(p_i\cdot q - i\eta_i\epsilon\right)
\left(-p_j\cdot q - i\eta_j\epsilon\right)}
\ee
where $i,j$ run over all the external legs.

Consider the virtual communication of $N$ particles, $m$ of which 
are soft photons and $n$ of which are soft Stuckelberg particles. 
We sum over all permutations and dividing by a factor $m!n!$ to 
account for spurious sums over identical soft photon and 
Stuckelberg particles and 
then sum over all $m$ and $n$ that add up to $N$. This gives 
the following contribution to the $S$-matrix,
\be
\frac{1}{N!}\sum_{m+n = N}\frac{N!}{m!n!}A_{\g}(q)^m A_{S}(q)^n = 
\frac{1}{N!}\left(A_{\gamma}(q) + A_{S}(q)\right)^N.
\ee

We pause here to note that in the photon soft factor in Eq.(3.3), 
the term arising from the mass correction in the propagator has the 
same form as the soft factor in Eq.(3.4). This term, proportional 
to $m_\gamma^{-2}$, diverges in the massless limit. It can be seen 
immediately that the condition for finiteness takes the form,
\be
\sum_{ij}\eta_i\eta_je_ie_j = \left(\sum_{i}\eta_i e_i\right)^2 = 0
\ee

\noindent which is simply charge conservation arising from global 
gauge invariance. On account of this, we drop 
this term henceforth, noting charge conservation as arising from 
global $U(1)$ invariance in the Stuckelberg QED. 
Now we may proceed to study the 
modifications experienced by the soft theorems due to the 
inclusion of mass coming from the denominators.

Summing $N$ from $0~to~\infty$, we obtain a correction to the 
$S$-matrix in the form of an exponential. Only the real part of the 
term in the exponential contributes to lifetimes with the mass-shell 
condition imposed by the propagator. It is given by,

\be
-\int_{E = m_\gamma}^{E = \Lambda}\frac{1}{2(2\pi)^3E}
\sum_{ij}\frac{e_ie_j\eta_i\eta_jp_i\cdot p_j}
{\left(p_i\cdot q \right)\left(p_j\cdot q \right)}d^3 q.
\ee
Now, employing the substitution $|\mathbf{p}| = m_\gamma \sinh(x)$, 
the integral may be rewritten as,

\be
-\int_{x=0}^{x=\xi_\Lambda}\frac{1}{2(2\pi)^3}
\sum_{ij}\frac{e_ie_j\eta_i\eta_jp_i\cdot p_j}
{\left(E_i \coth(x) - \mathbf{p}_i\cdot \hat{\mathbf{q}} \right)
\left(E_j \coth(x) - \mathbf{p}_j\cdot \hat{\mathbf{q}} \right)}dx 
d\Omega
\ee

\noindent where $\xi_{\Lambda} = \cosh^{-1}(\frac{\Lambda}{m_\g})$ 
and the solid angle integral is carried out over all directions of 
$\hat{\mathbf{q}}$.

For future reference, we define the following function,
\be
N\left(x, \hat{\mathbf{q}}\right) = \frac{1}{2(2\pi)^3}
\sum_{ij}\frac{e_ie_j\eta_i\eta_jp_i\cdot p_j}
{\left(E_i \coth(x) - \mathbf{p}_i\cdot \hat{\mathbf{q}} \right)
\left(E_j \coth(x) - \mathbf{p}_j\cdot \hat{\mathbf{q}} \right)}.
\ee
To compute the effects of the emissions of soft photons and 
Stuckelberg particles, the calculation involved is entirely 
analogous to that carried out by Weinberg\cite{Wein1,Wein2}. 
The factorization employed is identical to the one shown in the 
virtual case. We will not repeat the calculation here,  
as the procedure carried out in the same way. 
We may note however that the polarization sum in the massive case 
supplies a corrected propagator,
\be
\sum_\l \varepsilon(q,\lambda)_{\mu}\varepsilon^*(q,\lambda)_{\nu} 
~=~g_{\mu\nu}~+~\frac{q_\mu q_\nu}{m_\g^2}.
\ee
As mentioned earlier, the terms dependent on $m_\g^{-2}$ drop out 
of the correction terms, due to the global conservation of charge 
in the theory. This means that the terms that diverge as the mass 
of the photon vanishes, drop out of the calculation, the second 
term coming from the photon propagator and the Stuckelberg field.

It is worthwhile to note here that the effects of the Stuckelberg 
field are not at all observed in cross sections and decay rates 
insofar as an explicit interaction is concerned. Its existence is 
obtained from the introduction of the photon mass while preserving 
charge conservation but it is otherwise undetectable as a soft 
particle. The production of longitudinal/Stueckeberg particles are 
completely suppressed.

Having made these points, we may now note the correction to the 
transition rate that is effected on account of real emissions,
\be
\Gamma_{\a\b} = F(\Delta E)\Gamma^{0}_{\a\b}
\ee
where $\Delta E$ is the energy resolution of the theory and the 
function $F$ is defined by the transcendental expression,
\be
\begin{aligned}
&F(\Delta E) = \\ &\frac{1}{\pi}\int_{-\infty}^{\infty}\frac{\sin(\sigma \Delta E)}{\sigma}\exp\left(\int _{x=0}^{x =\xi_{\Delta E}}N(x,\hat{\mathbf{q}})e^{i\sigma m_{\gamma}\cosh(x)}dxd\Omega\right)d\sigma.
\end{aligned}
\ee
Here, $\xi_{\Delta E} = \cosh^{-1}(\frac{\Delta E}{m_\gamma})$.

We may recall that the inclusion of virtual processes entailed 
multiplication by a factor,
\be
\exp\left(-\int _{x=0}^{x =\xi_{\Lambda}}N(x,\hat{\mathbf{q}})dx
d\Omega\right).
\ee

Finally then, we may include both virtual and real processes. 
This means multiplying the transition rate by the following 
function,
\be
\exp\left(\int _{x=\xi_{\Lambda}}^{x=\xi_{\Delta E}}
N(x,\hat{\mathbf{q}})dx~d\Omega\right)G(\Delta E)
\ee
where,
\be
\begin{aligned}
&G(\Delta E)=\\
&\frac{1}{\pi}\int_{-\infty}^{\infty}\frac{\sin\sigma}{\sigma}
\exp\left(\int _{x=0}^{x =\xi_{\Delta E}}N(x,\hat{\mathbf{q}})
\left(e^{i\sigma \frac{m_{\gamma}}{\Delta E}\cosh(x)} - 1\right))
dx~d\Omega\right)~d\sigma.
\end{aligned}
\ee

At this point, it must be noted that a naive limit of 
$m_\g~\rightarrow ~0$ cannot be taken, since the integral has been 
regularized effectively, taking such a simple limit supplies the 
incorrect answer. One must instead replace $x$ with $E$ and then 
appeal to the smooth limit of vanishing photon mass. When one does 
this, the large values of $\Lambda/m_\g$ and $\Delta E/m_\g$ 
effectively impose $\coth(x)\sim 1$. Then, the first exponential 
factor reduces to Weinberg's energy dependent factor in eq. $2.51$ of Weiberg's 1965 calculation \cite{Wein1} and the second factor $G$ reduces to the soft function.

\section{\bf Modified Lienard Wiechert potentials 
and antipodal matching}
In the presence of a photon mass, modifications to the classical 
formulae of Lienard and Wiechert can be obtained. As was noted in 
\cite{Strom1}, there are discontinuities in the Lienard-Wiechert 
potentials at infinity, and it would be interesting to see if they 
indeed arise in our theory as well if the massless limit is taken.

As we shall see, the situation is not quite as simple as taking a 
naive massless limit. A suitable regulating procedure must be 
followed, and only under suitable relations being obeyed between 
the mass bound and the radius of the sphere taken formally at 
infinity do the discontinuities arise.

Having said this, we may actually calculate the potentials 
for a massive photon. For a charged particle at rest, the potential 
takes the form,
\be
A_t = -e^2Q\frac{e^{-m_\g~r}}{4\pi~r}.
\ee
The field strength then takes the form,
\be
F_{rt} = \frac{e^2Q}{4\pi}\frac{\left(m_\gamma r + 1\right)}{r^2}
e^{-m_\g~r}.
\ee
Now, in order to obtain the electric field due to the electron 
moving at constant velocity, we have to effect a Lorentz boost. 
Analogous to the field in eq $2.1.8$ in \cite{Strom1}, 
we obtain the following formula,

\be
F_{rt} = \frac{e^2Q}{4\pi}
\frac{\left(m_\g~R(\overrightarrow{\b}) + 1\right)}
{R(\overrightarrow{\b})^3}e^{-m_\g~R(\overrightarrow{\b})}
\g(r - t\hat{x}\cdot \overrightarrow{\b})
\ee
where $\g$ is the Lorentz factor, $\overrightarrow{\b}$ is the 
rapidity and,
\be
R(\overrightarrow{\b}) = |\g^2(t - r\hat{x}\cdot 
\overrightarrow{\b})^2 ~-~ t^2 + r^2|^{1/2}.
\ee
The derivation is almost 
similar to the standard one. We may now look at the formula in 
retarded coordinates $u~=~t~-~r$ (the advanced coordinate treatment 
is analogous) and take the limit $r\rightarrow\infty$. 
This supplies the following equation,
\be
\lim_{r,t\rightarrow\infty, u = const}F_{rt} = \frac{e^2 Q}{4\pi r^2} 
\frac{\alpha  + 1}{\g^2(1 - \hat{x}\cdot\overrightarrow{\b})^2}
e^{-\a}
\ee
where,
\be
\a~=~\g(1 - \hat{x}\cdot\overrightarrow{\b})m_{\g}r.
\ee
It is here that the precise relationship between the photon mass 
and the regulating procedure becomes clear. If $\a$ is taken to be 
unbounded, the expression is rapidly decreasing and vanishes 
at infinity. If however $\a$ is some positive number, 
the discontinuity manifests again.

The source of the discontinuity is on account of the fact that the 
order of the limits have bearing on the final answer. If the limit 
of large $r$ is taken first, the expression vanishes and no 
discontinuity is observed. If however, the limit of vanishing mass 
is first assumed, the answer reduces to the classical expression, 
and the discontinuity is recovered along with the antipodal 
matching condition.

These two limits correspond to leaving $\a$ unbounded and fixing $\a$
to be finite respectively. Note that we are ultimately 
interested in the limit,
\be
\a<< 1
\ee
which implies,
\be
m_\g<< \frac{1}{r}.
\ee
The current bounds on the mass of the photon will be consistent 
with this expectation. 

\section{\bf Conclusions}
In this note,we revisited the question of infrared divergence in 
QED, soft photons and asymptotic symmetries. This we approach in a
gauge invariant way in massive photon theory. This removes 
the discontinuity in the degrees of freedom as we take the limit 
of photon mass to zero. This was done using Stuckelberg theory 
which is the phase of the gauge field. We have a current limit of 
photon mass being less than $10^{-18}~eV$. Such a mass even though 
extremely small introduces extra degrees of freedom. This extra 
degree of freedom is coupled to the matter with very little strength 
and does not affect any of the known results of QED. Such a question
was posed by Schrodinger \cite{Erwin}  in relation to blackbody
radiation. But the field is essential for the asymptotic region 
as well as providing new Hilbert space in which QED will be 
free of infrared divergence along the lines of Faddeev and
Kulish \cite{faddeev} coherent states. Using this we can define 
gauge in variant matter field operator as $\tilde{\varphi}~=~
e^{-i~S}\varphi$. This will define the asymptotic states of the 
theory. In the limit of mass to zero carefully, the limit will still
provide $|in>$ and $|out>$ states. This can also be compared 
with the formulation of Bagan et al\cite{bagan}. As pointed out in 
the Introduction this analysis with exact massless QED is beset 
with problems like Lorentz symmetry violation\cite{BALQED, FROHLICH}.  
The analysis of limit of massive QED to massless is better 
understood when we use lightcone quantisation. Such an analysis 
has been done and will be presented elsewhere\cite{TRG1}. 
We can also consider 2+1 dimensional QED where 
there are novel mechanisms
of generations mass are available in addition to Stuckelberg
theory like Chern Simons term. The role of edge states in such a 
program playing the role of a Stuckelberg field is interesting
which again will be presented soon \cite{TRG}.
The lesson we learn from these studies is that the gauge 
non-invariant longitudinal component or 
Stueckelberg field can be made to be absent 
in the bulk inside a sphere of radius $\propto {\frac{1}{m}}$ the 
compton wavelength, but the generalised Robin conditions 
will provide localised states at the boundary. In the limit 
of $m ~\longrightarrow ~0$ they survive as zero energy modes 
and provide the required changes in the Hilbert space as well as 
contribute to asymptotic symmetries. This requires subtle 
conditions on the asymptotic behaviour along with the radius 
reaching infinity. 

One can ask whether the extra degree of freedom or Stuckelberg 
field has any role to play. Since the electromagnetic interaction 
is completely suppressed it can have only gravitational interaction.
It can play the part of dark matter component. Whether it 
is a substantial component or gets decoupled during evolution of the 
universe will be determined by the relic density. Very low 
mass of the photon can naively make it impossible to survive with 
appropriate densities. But recent analysis provide novel
BEC condensate being formed to facilitate the role 
as fuzzy dark matter candidates. This will be 
presented elsewhere. In this connection it is worth pointing 
out the possibility of ultralight dark matter candidates have 
been explored \cite{murphy,witten} .

\noindent{\bf Acknowledgements} TRG would like to thank Alok Laddha,
Ashoke Sen, A P Balachandran, Amitbha Lahiri and Amitabha Virmani
for discussions.


\begin{thebibliography}{}
\bibitem{PDB} 
Review of Particle Physics,
Particle Date Group, Phys Rev D 98, 030001 (2018)

\bibitem{Erwin}
L. Bass and E. Schrodinger, Proc. Royal Society of London. 
Mathematical and Physical Sciences, 232, 1188 (1955)

\bibitem{Wein1}
Steven Weinberg;
Infrared Photons and Gravitons;
Phys. Rev. 140, B516 (1965).

\bibitem{Wein3}
Steven Weinberg;
Photons and Gravitons in S-Matrix Theory: Derivation of Charge Conservation and Equality of Gravitational and Inertial Mass
Phys. Rev. 135, B1049.

\bibitem{Strom1}
Andrew Strominger;
Lectures on the Infrared Structure of Gravity and Gauge Theory;
arXiv:1703.05448

\bibitem{Strom2}
Andrew Strominger, Alexander Zhiboedov;
Gravitational Memory, BMS Supertranslations and Soft Theorems,
arXiv:1411.5745, J. High Energ. Phys. (2016) 2016: 86.

\bibitem{Strom3}
He, T., Lysov, V., Mitra, P. et al.; J. High Energ. Phys. (2015) 2015: 151.

\bibitem{Strom4}
He, T., Mitra, P., Porfyriadis, A.P. et al.; J. High Energ. Phys. (2014) 2014: 112.

\bibitem{ALOK}
M Campiglia, Alok Laddha, 
Asymptotic symmetries of QED and Weinberg's 
soft photon theorems, arXiv hep-th/1505.05346i,
J. High Energ. Phys. (2015) 2015: 115.
; 
Sub-subleading soft gravitons: New symmetries of quantum gravity?, Physics Letters B, Volume 764, 2017, 218-221,
arXiv gr-qc/1605.09094. 

\bibitem{BALQED} 
A P Balachandran and S Vaidya, 
Spontaneous Lorentz violation in Gauge theories, arXiv hep-th/1302.3406v2, Eur. Phys. J. Plus (2013) 128: 118.;
A P Balachandran, S Kurkcuoglu, A R de Queiroz and S Vaidya, 
Spontaneous Lorentz Violation: The Case of Infrared QED,
Eur. Phys. J. C 75 (2015) 89.

\bibitem{BALQCD}A P Balachandran, QCD Breaks Lorentz Invariance 
and Colour, Mod Phys Lett. A 31 (2016) 1650060

\bibitem{Photon}
Luca Bonetti, John Ellis, Nikolaos E.Mavromatos, Alexander 
S.Sakharov, Edward K.Sarkisyan-Grinbaumgh, 
Alessandro D.A.M.Spallicci;Photon mass limits from fast 
radio bursts, Physics Letters B757 (2016),548

\bibitem{Photon2}
A S Goldhaber and M M Nieto, arXiv hep-ph/0809.1003v5. 
Rev.Mod.Phys. 82 (2010) 939-979

\bibitem{Ruegg}
Henri Ruegg, Marti Ruiz-Altaba;
The Stuckelberg Field,
Int.J.Mod.Phys.A19:3265-3348,2004.

\bibitem{Wein2}
Steven Weinberg; (1995). The Quantum Theory of Fields, Vol 1,
Cambridge University Press. doi:10.1017/CBO9781139644167

\bibitem{faddeev}
P. P. Kulish and L. D. Faddeev, Theor. Math. Phys. 4, 745 (1970).

\bibitem{bagan} 
Emili Bagan, Martin Lavelle and David McMullan,
arXive hep-ph/9909257v2 (2000),
Annals Phys. 282 (2000) 471-502

\bibitem{FROHLICH}
J Frohlich, G Morchio and F Strocchi, 
Phys. Lett. B 89, 61 (1979).

\bibitem{TRG1}
T R Govindarajan, Jai More, P Ramadevi and V Ravindran, 
To be submitted.

\bibitem{TRG}
T R Govindarajan and Rakesh Tibrewala, To be submitted

\bibitem{murphy} H Davoudiasl and C W Murphy,
Phys.Rev.Lett. 118 (2017) 141801; arXiv:1701.01136
 
\bibitem{witten}
Lam Hui, J P Ostriker, S Tremaine and E Witten,  
Ultralight scalars as cosmological dark matter,
arXiv astro-ph/1610.08297v2, Phys. Rev. D 95, 043541 (2017); 

Jae-Weon Lee, Brief History of Ultra-light Scalar Dark Matter Models,
arXiv astro-ph/1704.05057v1, EPJ Web of Conferences 168, 06005 (2018).


\end{thebibliography}
\end{document}